\documentclass[aps,pra,showpacs,superscriptaddress,notitlepage,twocolumn]{revtex4-1}

\usepackage{mathrsfs}
\usepackage{amsfonts}
\usepackage{amssymb}
\usepackage{amsmath}
\usepackage{bm}
\usepackage{graphicx}
\usepackage{sidecap}
\usepackage{color}
\usepackage[colorlinks=true,citecolor=blue,linkcolor=magenta]{hyperref}
\usepackage{subeqnarray}
\usepackage{verbatim}
\usepackage{ulem}
\usepackage{cases}
\usepackage{cancel}

\newcommand{\ket}[1]{\vert #1 \rangle}
\newcommand{\bra}[1]{\langle #1 \vert}

\begin{document}
\title{Protection of Logical Qubits via Optimal State Transfers}
\author{Jiang Zhang}
\affiliation{Interdisciplinary Center of Quantum Information and Zhejiang Province Key Laboratory of Quantum Technology and Device, Department of Physics and State Key Laboratory of Modern Optical Instrumentation, Zhejiang University, Hangzhou 310027, China}
\affiliation{State Key Laboratory of Low-Dimensional Quantum Physics and Department of Physics, Tsinghua University, Beijing 100084, China}
\author{Zheng-Yang Zhou}
\affiliation{Theoretical Quantum Physics Laboratory, RIKEN Cluster for Pioneering Research, Wako-shi, Saitama 351-0198, Japan}
\author{Lian-Ao Wu}
\affiliation{Department of Theoretical Physics and History of Science, University of the Basque Country, EHU, UPV, P.O. Box 644, 48080 Bilbao, Spain}
\affiliation{Ikerbasque, Basque Foundation for Science, 48011 Bilbao, Spain}
\author{J. Q. You}
\thanks{jqyou@zju.edu.cn}
\affiliation{Interdisciplinary Center of Quantum Information and Zhejiang Province Key Laboratory of Quantum Technology and Device, Department of Physics and State Key Laboratory of Modern Optical Instrumentation, Zhejiang University, Hangzhou 310027, China}

\begin{abstract}
Dynamical decoupling can enforce a symmetry on the dynamics of an open quantum system. Here we develop an efficient dynamical-decoupling-based strategy to create the decoherence-free subspaces (DFSs) for a set of qubits by optimally transferring the states of these qubits.
We design to transfer the state of each qubit to all of the other qubits in an optimal and efficient manner, so as to produce an effective collective-type noise for all qubits. This collective {\it pseudo-noise} is essentially derived from arbitrarily independent {\it real} baths and needs only nearest-neighbor state transfers for its implementation.
Moreover, our scheme requires only $N$ steps to produce the effective collective-type noise for a system of $N$ qubits. It provides an experimentally feasible and efficient approach to achieving the DFSs for encoding the protected logical qubits.
\end{abstract}

\maketitle

\section{Introduction}
In quantum information processing, symmetry plays a central role in protecting qubits from quantum errors~\cite{di,ekert} (see, e.g., quantum error correction~\cite{shor,steane96,laflamme96,terhal} and error suppression using dynamical decoupling (DD)~\cite{viola99,duan99pla,zanardi99pla,viola00,kh05,uh07,du09,kh10,zhang15}).
When a set of qubits coupled to their environment with a permutation symmetry,
the Hilbert space spanned by these physical qubits can support decoherence-free subspaces (DFSs) for encoding the protected logical qubits~\cite{duan97prl,zanardi97,lider98,duan98pra1,zanardi98,KLV,DFS-exp1,DFS-exp2,NS-exp}.
Furthermore, it has been shown that the two-body Heisenberg interactions in solid-state qubits~\cite{loss98,kane98,vrijen00,antonio13} can be used to realize the logical operations in a DFS~\cite{bacon00,di00}, and the DFS theory is compatible with other quantum computation (QC) approaches, such as topological QC~\cite{zanardi03} and holonomic QC~\cite{wu05,ore09,xu12,zhang14}.

The main obstacle in implementing QC with DFSs is that the required permutation symmetry in the interaction between qubits and their environments is hardly available in nature. Consequently, a protocol to artificially symmetrize the interaction was developed via DD by using the permutation group~\cite{zanardi99pla,viola00}.
Also, it was pointed out in Ref.~\cite{zanardi99pla} that using the cyclic group{\bf--}a subgroup of the permutation group{\bf--}can achieve this symmetrization as well. However, these proposals are based on multi-qubit permutations which cannot be constructed directly. This hinders their applications to the realistic systems of many qubits.
On the other hand, it was also constructively shown that the controllable Heisenberg interaction can drive explicit sequences of DD pulses to create the conditions allowing for the existence of DFSs~\cite{wu02} for two to four qubits.
This inspires us to separate the multi-qubit permutations into two-qubit state transfers.

In this work, we develop an efficient DD-based method to generate DFSs for a multi-qubit system.
For an array of $N$ physical qubits interacting with the environment via arbitrarily independent couplings, our method is to transfer the state of each physical qubit to all the other physical qubits {\it once and only once}, so as to accomplish a state-transfer cycle. Here only $N$ steps are needed for this cycle. After the cycle, the state of each physical qubit has gone through all physical qubits and experienced, in the same time interval, the noise affecting each qubit.
The total effect of this cycle is to sum up all noises and then apply to every single qubit. Hence the state of each physical qubit  {\it effectively} suffers the same noise, i.e., an effective {\it collective-type} environment is produced via the state-transfer cycle. In such a way, a collective pseudo-bath is created from the realistic bath modelled by independent errors.
Also, an example in Ref.~\cite{zanardi99pla} shows that for exactly the same interaction Hamiltonian considered here, only cyclic permutations, instead of the full permutation group~\cite{viola00}, are needed. Using the method in the present work, we can derive that the scheme in Ref.~\cite{zanardi99pla} requires $(N-1)^2$ steps for an $N$-qubit system to implement the cyclic permutations via the two-qubit operations, in contrast to the $N$ steps needed in our approach. Our scheme not only makes the strategy in Refs.~\cite{zanardi99pla,viola00} experimentally feasible by separating the multi-qubit permutations into experimentally realizable two-qubit state transfers, but also has distinct superiority for the many-qubit systems owing to its polynomial speedup over the previous approaches.
This makes it possible to create a higher-dimensional DFS to encode more protected logical qubits for fault-tolerant QC.

In practice, the control Hamiltonians generating the decoupling operators cannot be too strong and the time interval between two adjoining pluses is finite. Therefore, higher-order errors arise in the effective Hamiltonian. For a sufficiently long time, the accumulation of the errors may have appreciable effects on the generated DFSs.
To show the implementation of our scheme in a long-time period, we consider higher-order errors in both periodic and concatenated DD approaches and derive a condition under which the concatenated DD is superior to the periodic DD.
Moreover, to check the validity of our scheme, we further perform numerical simulations to demonstrate the fidelity between the initial state stored in a DFS and the state obtained from the quantum dynamics that includes the DD process. The results show that for the control Hamiltonians with practical coupling strengths, the fidelities in, e.g., two- and four-qubit systems are greater than 0.99 and 0.95, respectively, indicating that our scheme is implementable for realistic systems. Also, an explicit time interval at which the concatenated DD performs better than the periodic DD is numerically found in the four-qubit system.

\section{Converting independent realistic baths to a collective pseudo-bath}
We study an open quantum system described by the total Hamiltonian $H_0=H_S +H_B+ H_{SB}$, where $H_S$ is the Hamiltonian of the considered system consisting of $N$ physical qubits, $H_B$ is the Hamiltonian of the environment, and $H_{SB}$ is the interaction between them. As a general case, we consider an environment with $N$ independent baths, each coupling to a qubit, and $H_{SB}$ has the form
\begin{equation}
  H_{SB}= \sum_{i,\alpha} \sigma_{\alpha}^{(i)}\otimes B_{\alpha}^{(i)},
\label{H-sb}
\end{equation}
where $\sigma_{\alpha}^{(i)}$ ($\alpha=x,y,z$) are Pauli operators of the $i$th qubit and $B_\alpha^{(i)}$ are the related operators of the $i$th bath.

We add a control Hamiltonian $H_1(t)$ to $H_0$, so as to generate a periodic evolution operator $V(t)$ on the qubits,
\begin{equation}\label{u1}
V(t)\equiv {\cal T}\exp[-i\int_0^t\mathrm{d}uH_1(u)]=V(t+T_c),
\end{equation}
where $T_c$ is the period and $\cal{T}$ denotes the time ordering operator.
In the interaction picture associated with $H_1(t)$, $H_0$ becomes
\begin{equation}\label{h0}
\tilde{H}_0(t)=V^\dag(t) H_0 V(t).
\end{equation}
Because $V(T_c)$ is an identity operator on the qubits, at $T=mT_c$ with $m$ being an integer, the evolution operator of the total system in the Schr\"{o}dinger picture can be expressed, using the Floquet-Magnus expansion~\cite{blanes09}, as
\begin{equation}
U(T)={\cal T}\exp[-i\int_0^{T}\mathrm{d}u\tilde{H}_0(u)]=e^{-i(\bar{H}_0^{(0)}+\bar{H}_0^{(1)}+\cdots)T},
\end{equation}
where $\sum_i\bar{H}_0^{(i)}$ is the Magnus series, and the zeroth-order Hamiltonian can be written as
$\bar{H}^{(0)}_0=(1/T_c)\int_0^{T_c}\mathrm{d}u\tilde{H}_0(u)$.
Here we focus on the limit of fast control, where the contributions higher than the zeroth order are neglected~\cite{zanardi99pla,viola00,blanes09}, i.e.,
\begin{equation}\label{u}
U(T)\approx e^{-i\bar{H}^{(0)}_0T}=e^{-iH_0^{\mathrm{eff}}(T)T},
\end{equation}
where the effective Hamiltonian of the system at the time instants $T=mT_c$ is given by
\begin{equation}\label{he}
  H_0^{\mathrm{eff}}(T)=\frac{1}{T_c}\int_0^{T_c}\mathrm{d}uV^\dagger(u)H_0V(u).
\end{equation}

\begin{figure}[tbp]
     \centering
     \includegraphics[width=.48\textwidth]{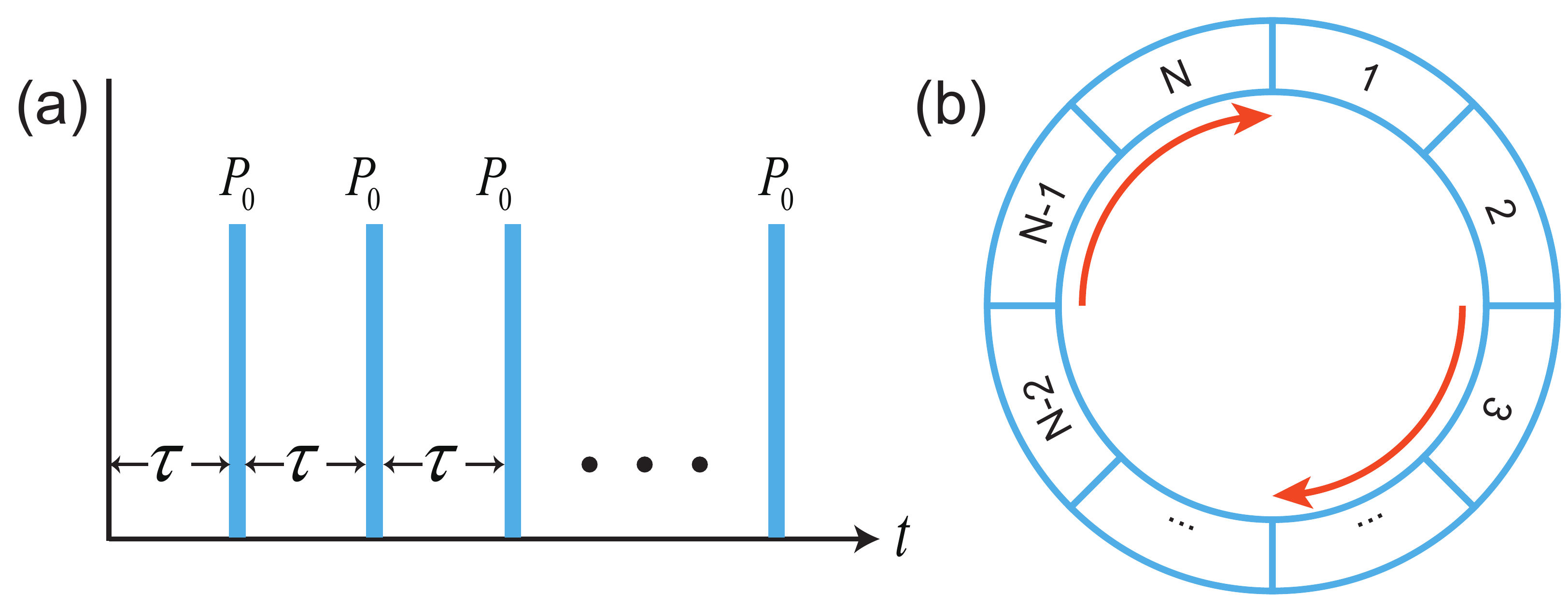}
     \caption{(a) Control sequence in $V(t)$. First let the qubits interact with their environments and evolve for a time interval $\tau=T_c/N$. Then, apply a state-transfer operator $P_0$ to the qubits. Repeat this procedure for $N-1$ times to implement  $V(t)$. (b) Schematic diagram of our method. Application of $P_0$ shifts the qubit state in each qubit to the qubit having one site forward around the ring clockwise. By performing $P_0$ for $N-1$ times, each qubit state will then go through all of the other qubits and suffer their noises. Application of $P_0$ once more will result in each qubit state returning to the original qubit.}
     \label{fig1}
\end{figure}

To obtain the needed effective Hamiltonian, it is essential to have a proper $V(t)$. Here we design it as
a piecewise controller [cf. Fig.~\ref{fig1}(a)],
\begin{equation}\label{uc}
  V(t)=\left\{
             \begin{array}{ll}
               I^{\otimes N}, & \hbox{$0\leq t<T_c/N$;} \\
               P_0, & \hbox{$T_c/N\leq t<2T_c/N$;} \\
               P^2_0, & \hbox{$2T_c/N\leq t<3T_c/N$;} \\
               P^3_0, & \hbox{$3T_c/N\leq t<4T_c/N$;} \\
               \cdots, & \hbox{$\cdots\cdots$;} \\
               P^{N-1}_0, & \hbox{$(N-1)T_c/N\leq t<T_c$,}
             \end{array}
           \right.
\end{equation}
where $I^{\otimes N}$ is an identity operator on all the $N$ qubits,
$P_0$ is a state-transfer operator (i.e., a permutation operator) denoted by the cyclic notation  $(1,2,3,\cdots,N)$ which means transferring the state of the first qubit to the second qubit, the state of the second qubit to the third qubit, and so on.
Explicitly, the application of $P_0$ on an $N$-qubit state can be written as $P_0\otimes_{i=1}^N\ket{\psi_i}=\otimes_{i=1}^N\ket{\psi_{P_0(i)}}$, with $P_0(i)=i+1$ and the periodic boundary condition $N+1=1$.
The operator $P_0^i$ ($i=1,2,3,\cdots,N-1$) denotes application of the state-transfer operation for $i$ times.
Since $P_0^\dagger=(N,N-1,N-2,\cdots,1)$, we have
\begin{align}\label{}
  P_0^\dagger \sigma_\alpha^{(i)} P_0 & =P_0^\dagger(\ket{+}_\alpha^{(i)}\bra{+}-\ket{-}_\alpha^{(i)}\bra{-})P_0  \nonumber\\
   &= \ket{+}_\alpha^{(i-1)}\bra{+}-\ket{-}_\alpha^{(i-1)}\bra{-} \nonumber \\
   &=\sigma_\alpha^{(i-1)},
\end{align}
where $\ket{\pm}_\alpha^{(i)}$ are the eigenstates of $\sigma_\alpha^{(i)}$ corresponding to the eigenvalues $\pm 1$ and the periodic boundary condition is used.
By substituting $V(t)$ in Eq.~(\ref{uc}) into Eq.~(\ref{he}), the interaction Hamiltonian $H_{SB}$ is converted to
\begin{align}\label{hsbe}
  H_{SB}^{\mathrm{eff}}(T)  = &\frac{1}{T_c}\left[\frac{T_c}{N}H_{SB}+\frac{T_c}{N}P_0^\dag H_{SB}P_0+\frac{T_c}{N}(P_0^2)^\dag H_{SB}P_0^2 \right. \nonumber \\
  &\left.+\cdots+\frac{T_c}{N}(P_0^{N-1})^\dag H_{SB}P_0^{N-1}\right] \nonumber \\
   =& \frac{1}{N}\sum_{i,\alpha}\left[\sigma_\alpha^{(i)}\otimes B_\alpha^{(i)}
   +\sigma_\alpha^{(i-1)}\otimes B_\alpha^{(i)}\right.\nonumber \\
   &\left.+\sigma_\alpha^{(i-2)}\otimes B_\alpha^{(i)}+\cdots+\sigma_\alpha^{(i-N+1)}\otimes B_\alpha^{(i)}\right]\nonumber \\
   =& \frac{1}{N}\sum_{i,\alpha}\left[\sigma_\alpha^{(1)}+\sigma_\alpha^{(2)}+\sigma_\alpha^{(3)}+\cdots+\sigma_\alpha^{(N)}\right]\otimes B_\alpha^{(i)} \nonumber \\
   =& \sum_{\alpha}S_\alpha\otimes B_\alpha^{(\rm env)},
\end{align}
where $S_\alpha\equiv\sum_{i=1}^{N}\sigma_\alpha^{(i)}$ are collective operators of the system and
$B^{(\rm env)}_\alpha\equiv\frac{1}{N}\sum_iB_\alpha^{(i)}$ are collective operators of the environment.
Owing to $V(t)$, the {\it independent}-type interaction Hamiltonian $H_{SB}$ is converted to the {\it collective}-type interaction Hamiltonian $H_{SB}^{\mathrm{eff}}(T)$ at the time instants $T=mT_c$, with all qubits effectively coupled to the {\it same} pseudo-bath~\cite{zanardi97,KLV}.

The physical mechanism of our method can be clearly illustrated. At each time instant $t=mT_c/N$, we apply a $P_0$ on all qubits to transfer the state of the $i$th qubit to the $(i+1)$th qubit; after applying $P_0$ for $(N-1)$ times, each qubit state has gone through all qubits and suffered, in the same time interval, all the individual noises acting on these qubits.
This accomplishes a state-transfer cycle that amounts to summing all noises and then applying the summed noise to every qubit equally. Thus, an {\it effective} collective-type noise is created for the system. When performing $P_0$ one more time, each qubit state then returns to its original qubit.
Note that even though we use the periodic boundary condition, it is not necessary to place the qubits geometrically along a ring shown in Fig.~\ref{fig1}(b). However, a state transfer between the first and last qubits should be implementable, which is a basic requirement for the circuit QC model and has been demonstrated for, e.g., two remote superconducting qubits~\cite{ma07}.

For the controller $V(t)$ in Eq.~(\ref{uc}), the permutations $\{I^{\otimes N}, P_0, P_0^2, \cdots, P_0^{N-1}\}$ represent an explicit example of the cyclic group of $N$ objects which is an Abelian subgroup of the permutation group. Since the cyclic group also possesses the permutation symmetry, it can generate the same collective environment. Compared with use of the permutation group~\cite{viola00}, use of the cyclic group as the decoupling group reduces the number of the permutations required in the decoupling procedure from $N!$ to $N$.

All $S_\alpha$ in Eq.~(\ref{hsbe}) are total spin angular momentum operators acting on $N$ qubits. Their commutation relation indicates that they form an algebra isomorphic to sl(2).
The irreduciable representation $\mathcal{D}_j$ of sl(2) can be labeled by the total angular momentum eigenvalues $j$ with a dimension $d_j=2j+1$. Therefore, we have the following Clebsch-Gordan decomposition in terms of the $\mathcal{D}_j$~\cite{cornwell}:
$S_\alpha=\bigoplus_j n_j\mathcal{D}_j$,
where the integer $n_j$ is the multiplicity for $\mathcal{D}_j$ to occur in the solution of $S_\alpha$.
In particular, for $j=0$ ($N$ being even), $\mathcal{D}_0$ corresponds to the dark-state subspace of $S_\alpha$, i.e., any state $\ket{\psi}$ in $\mathcal{D}_0$ satisfies $S_\alpha\ket{\psi}=0$.
A quantum state prepared in such a subspace is not affected by $H_{SB}^{\mathrm{eff}}$ because $H_{SB}^{\mathrm{eff}}\ket{\psi}=0$. Therefore, $\mathcal{D}_0$ is a DFS which can be used to encode the {\it protected} logical qubits for fault-tolerant QC.
For example, when $N=2$, $\ket{\psi_1}=\frac{1}{\sqrt{2}}(\ket{01}-\ket{10})$ is such a decoherence-free quantum state.
When $N=4$, where $n_0=2$, there are two orthogonal decoherence-free states  $\ket{\psi_2}=\frac{1}{2}(\ket{0101}-\ket{1001}-\ket{0110}+\ket{1010})$ and $\ket{\psi_3}=\frac{1}{2\sqrt{3}}(2\ket{0011}-\ket{0101}-\ket{1001}-\ket{0110}-\ket{1010}+2\ket{1100})$.
Thus, a two-dimensional DFS is available for encoding a protected logical qubit.

\begin{figure}[tbp]
     \centering
     \includegraphics[width=.45\textwidth]{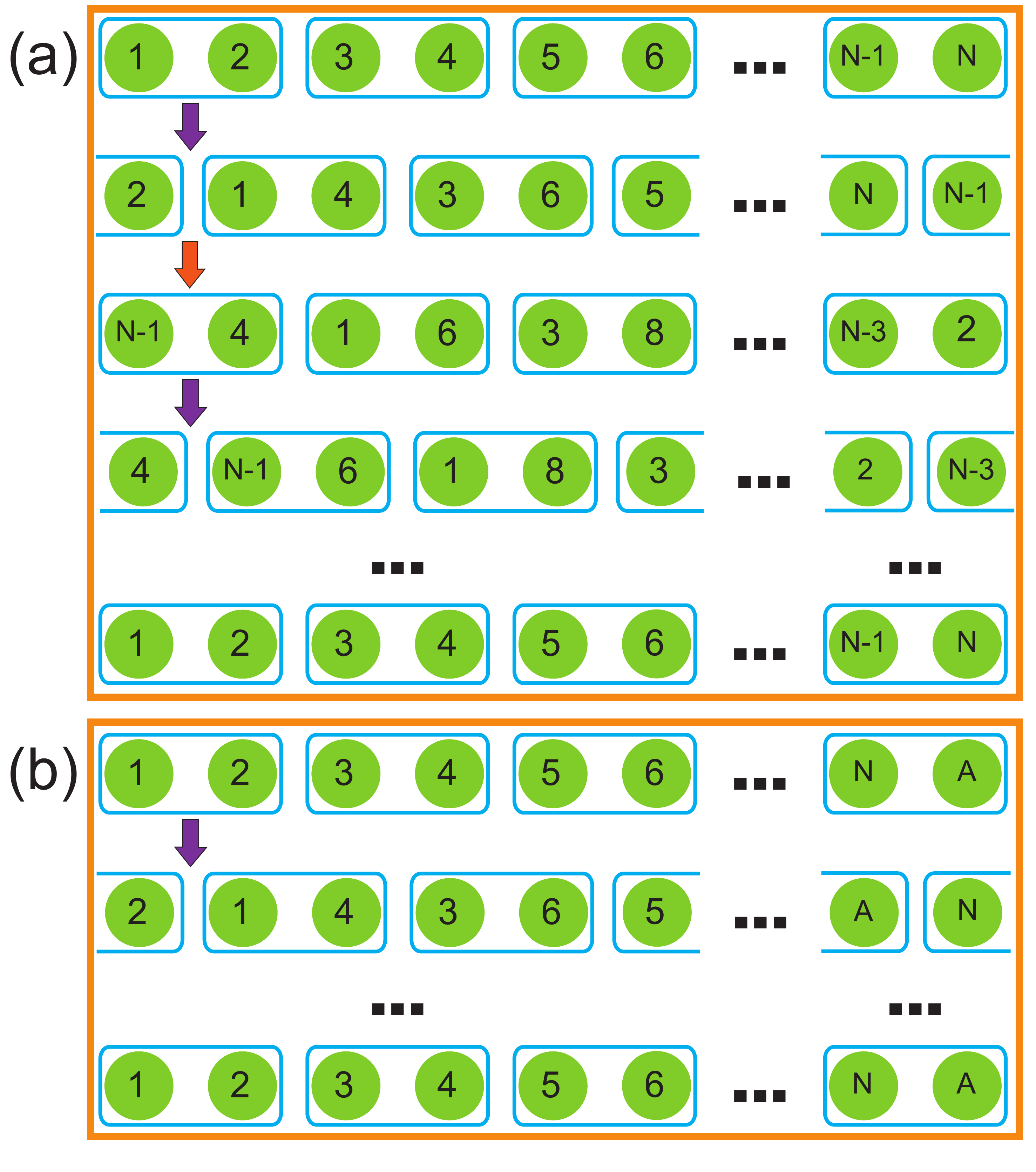}
     \caption{The procedure for achieving the state-transfer cycle. (a) The even-$N$ case. Each green solid circle represents a physical qubit and the qubit array is divided into pairs by the blue boxes. The numbers in the solid green circles denote  given qubit states. In the first row, the $i$th qubit originally has the state indexed by $i$.  When performing a state exchange between qubits in each box, the qubit states are transferred to the arrangement shown in the second row. Then, performing a state exchange between qubits in each box in the second row transfers the qubit states to the arrangement shown in the third row, and so on. After performing state exchanges for $N$ times, each qubit state will go through all of the other qubits and then return to the original qubit. This accomplishes the state-transfer cycle in the even-$N$ case. (b) The odd-$N$ case. An auxiliary qubit $A$ is added at the end of the qubit array and it is paired with the $N$th qubit. Similar to the case in (a), the state-transfer cycle is accomplished by performing the state exchange between qubits in each box $N+1$ times.}
     \label{fig2}
\end{figure}

\section{Realization of the controller $V(t)$ via optimal state transfers}
Equation (\ref{uc}) presents an efficient way to achieve the state-transfer cycle, but it cannot be directly realized because $P_0$ in $V(t)$ is a multi-qubit operation whose realization requires a multi-body interaction Hamiltonian. Alternatively, one may achieve $P_0$ by decomposing it into $N-1$ state exchanges: 
$P_0=E_{1,N}E_{1,N-1}\cdots E_{1,2}$, where each $E_{i,j}$ is a qubit-state exchange operator (i.e. a two-qubit swap gate) acting on the $i$th qubit and the $j$th qubit.
This requires a total number of $(N-1)^2$ steps to implement $V(t)$.
A given example (i.e., example 5) in Ref.~\cite{zanardi99pla} also shows that for exactly the same interaction Hamiltonian considered in Eq.~(\ref{H-sb}), the full permutation group ($S_n$) is not needed, but only cyclic permutations ($Z_n$) are required to generate the effective collective interaction Hamiltonian $H_{SB}^{\mathrm{eff}}$. Cyclic permutations are the equivalent of Fig.~\ref{fig1}(b), so $(N-1)^2$ steps are also required for the case in Ref.~\cite{zanardi99pla} when two-qubit permutations are used. However, it is still tedious when using the cyclic permutations to produce $H_{SB}^{\mathrm{eff}}$, because $(N-1)^2$ involves too many steps when the number $N$ of qubits in the system becomes large.
To solve this problem, instead of harnessing $P_0$, below we develop an efficient scheme to accomplish the required state-transfer cycle with only nearest-neighbor state transfers (see Fig. \ref{fig2}), in which the state-transfer cycle can be accomplished in either $N$ steps when $N$ is even or $N+1$ steps when $N$ is odd.

We first consider the even-$N$ case.  By dividing the qubit array into $N/2$ pairs [see the first row in Fig.~\ref{fig2}(a)], the first step in our state-transfer cycle is to perform $N/2$ state exchanges ($E_{12}$, $E_{34}$, and so on) at the same time, transferring the qubit states into the arrangement shown in the second row.
Here the qubit-state exchanges can be achieved using, e.g., the controllable nearest-neighbor Heisenberg interactions~\cite{duan03,tro08} $H_{I}=J(t)\sum_{\alpha=x,y,z}\sigma_\alpha^{(i)}\otimes\sigma_\alpha^{(i+1)}$, where $i=1,3,\cdots,\frac{N}{2}-1$, when the interactions are turned on for a time $T_0$ to have $\int_{0}^{T_0}J(t)\mathrm{d}t=\pi/4$. Similarly, these qubit-state exchanges can also be realized via the controllable nearest-neighbor $XY$-type interactions~\cite{schuch,ma07,Tanamoto,mar11} $H_{XY}=J(t)[\sigma_x^{(i)}\otimes\sigma_x^{(i+1)}+\sigma_y^{(i)}\otimes\sigma_y^{(i+1)}]$.
In the second step, we divide the qubit array into pairs again according to the blocks in the second row in Fig.~\ref{fig2}(a) and perform the corresponding qubit-state exchanges. Then, we implement the third step and so on.
From Fig. \ref{fig2}(a), we see that after each step, states of the odd-numbered qubits are shifted one site rightward while states of the even-numbered qubits are shifted one site leftward. Therefore, the state of each qubit will go through all of the other qubits in $N-1$ steps and then returns to its original qubit after implementing the $N$th step.

As in Fig.~\ref{fig1}(b), we can describe the above procedure using the permutation operators. For example, the first step contains qubit-state exchanges between qubits 1 and 2, qubits 3 and 4, and so on. Therefore, it can be denoted using the permutation operator as $P_1=E_{1,2}E_{3,4}\cdots E_{N-1,N}$. The qubit-state exchanges in the second step are between qubits 2 and 3, qubits 4 and 5, and so on. Accordingly, it can be denoted as $P_2=E_{N,1}E_{2,3}\cdots E_{N-2,N-1}$. From Fig.~\ref{fig2}(a), we can see that $P_1$ describes the qubit-state exchanges in all the odd-numbered steps and $P_2$ describes the qubit-state exchanges in all the even-numbered steps. Therefore, the controller $V(t)$ in Eq.~(\ref{uc}) can be {\it equivalently} written as
\begin{equation}\label{u1p}
  V(t)=\left\{
             \begin{array}{ll}
               I^{\otimes N}, & \hbox{$0\leq t<T_c/N$;} \\
               P_1, & \hbox{$T_c/N\leq t<2T_c/N$;} \\
               P_2P_1, & \hbox{$2T_c/N\leq t<3T_c/N$;} \\
               P_1P_2P_1, & \hbox{$3T_c/N\leq t<4T_c/N$;} \\
               (P_2P_1)^2, & \hbox{$4T_c/N\leq t<5T_c/N$;} \\
               \cdots, & \hbox{$\cdots\cdots$;} \\
               P_1(P_2P_1)^{\frac{N}{2}-1}, & \hbox{$(N-1)T_c/N\leq t<T_c$.}
             \end{array}
           \right.
\end{equation}
Substituting $V(t)$ in Eq.~(\ref{u1p}) into Eq.~(\ref{he}), we can verify that the above $V(t)$  produces the same effective collective Hamiltonian $H_{SB}^{\mathrm{eff}}(T)$ as $V(t)$ in Eq.~(\ref{uc}) at the time instants $T=mT_c$ (see Appendix~\ref{a}).
By adding an auxiliary qubit $A$ at the end of the qubit array, the above procedure can be directly generalized to the system with an odd number ($N$) of qubits. For such a case, the corresponding procedure is similar and the qubit-state transfer cycle can be realized in $N+1$ steps [see Fig.~\ref{fig2}(b)].

Finally, we explain why using the controller $V(t)$ in Eq.~(\ref{u1p}) to generate the collective interaction $H_{SB}^{\mathrm{eff}}$ is optimal. In our method, the state of each qubit goes through all the qubits and finally returns to its original one. For an array of $N$ qubits, this needs $N^2$ moves since the state of each qubit has $N$ moves. However, a permutation operator acting on $N$ qubits can at most produce $N$ moves, with each qubit having a move. Therefore, at least $N$ permutation operators are needed in a controller to accomplish the required $N^2$ moves. The controller $V(t)$ in Eq.~(\ref{u1p}) is exactly such a case, so it is {\it optimal} in generating $H_{SB}^{\mathrm{eff}}$.

\section{Higher-order errors and concatenated dynamical decoupling}
So far, our optimal DFS-generating scheme has only considered decoherence up to the first order in time.
In the ideal case, DFSs can be created via the periodic DD (i.e., applying our scheme periodically) \cite{viola99}.
However, in a practical case, higher-order errors may arise due to a finite time interval between two adjoining pluses.
In this section, we propose to use the concatenated DD \cite{kh05} to eliminate higher-order errors and calculate the exact forms of errors for both periodic and concatenated DDs. We obtain a condition for the concatenated DD to be superior to the periodic DD.

Below we consider the case where the pulses are realized instantaneously but the time interval $\tau$ between two adjoining pulses is constant. The corresponding DD procedure is as follows: Let the total system evolve for a time $\tau$ governed by the Hamiltonian $H_0$, and then apply the first decoupling operator $g_1$; let the system evolve for a second $\tau$, and apply the operator $g_2g_1^\dag$; let the system evolve for another $\tau$, and apply the operator $g_3g_2^\dag$; and so on.
The total evolution operator of the above procedure can be written as
\begin{equation}\label{u01}
  U_0(T)=\prod_{k=0}^{m-1} g_{k}^\dag\exp\{-iH_0\tau\}g_{k},
\end{equation}
where $g_0$ is the identity operator, $m$ is the number of decoupling operators, and $T=m\tau$.
When $\tau$ is small, we can use the Baker-Campbell-Hausdorff (BCH) formula to transform $U_0(T)$ to (see Appendix \ref{b})
\begin{equation}\label{u02}
  U_0(T)=\exp\{-i\tau\sum_{k=0}^{m-1}g_k^\dag H_0g_k-\frac{\tau^2}{2}\sum_{j>k}[g_j^\dag H_0g_j,g_k^\dag H_0g_k]\},
\end{equation}
up to the second order of time (i.e., $\tau^2$).
It is clear that the first-order contribution is the target effective Hamiltonian, but the second-order one is an error which could accumulate when the evolution time is sufficiently long.

To study the long-time effect of the second-order error, we first consider the periodic DD approach which performs the decoupling process successively.
For example, if the total evolution time is $nT$ ($n$ being an integer), we can separate it to $n$ equal parts, and in each part we implement the decoupling as described in Eq.~(\ref{u01}).
In such a way, the corresponding total evolution operator is
\begin{equation}\label{updd}
  U_p(nT)=\exp\{-in\tau\bar{H}_0-n\frac{\tau^2}{2}\sum_{j>k}[g_j^\dag H_0g_j,g_k^\dag H_0g_k]\},
\end{equation}
where $\bar{H}_0=\sum_{k=0}^{m-1}g_k^\dag H_0g_k$.
As in Eq.~(\ref{u02}), the first term in Eq.~(\ref{updd}) is the effective Hamiltonian that we design, while the second term is an error which is denoted as $E(\tau^2nH_p)$, with $H_p=\frac{1}{2}\sum_{j>k}[g_j^\dag H_0g_j,g_k^\dag H_0g_k]$.
It is clear that the error term emerging from the periodic DD accumulates linearly with the time.

In the concatenated DD approach, the decoupling process is implemented recursively. For example, when the total evolution time is $mT$ ($m$ being an integer), the decoupling process can be written as
\begin{equation}
  U_c(mT)=\prod_{k=0}^{m-1}g_k^\dag U_0(T)g_k.
\end{equation}
Using the BCH formula, we can rewrite $U_c(mT)$ as
\begin{eqnarray}\label{ucdd}
  U_c(mT)&=&\exp\{-im\tau\bar{H}_0-\tau^2\sum_{k=0}^{m-1}g_k^\dag H_pg_k \nonumber \\
         & &+i\frac{\tau^3}{4}\sum_{k=0}^{m-1}(m-2k-1)[\bar{H}_0,g_k^\dag H_pg_k]\}.~~~
\end{eqnarray}
For details of the derivation, see Appendix \ref{b}.
A significant merit of the concatenated DD is that it transforms the error Hamiltonian $H_p$ in the periodic DD to a harmless contribution since it has the same symmetry as $\bar{H}_0$, only leaving the $\tau^3$ term as a higher-order error.

If we denote the error term in Eq.~(\ref{ucdd}) as $E(\tau^3H_c)$, with $H_c=\frac{i}{4}\sum_{k=0}^{m-1}(m-2k-1)[\bar{H}_0,g_k^\dag H_pg_k]$, it is clear that we can have $E(\tau^3H_c)<E(\tau^2mH_p)$ for a sufficiently small $\tau$, indicating that the concatenated DD is better than the periodic DD in this small $\tau$ case.
However, in practice, the interval $\tau$ between two adjoining pulses cannot be too small.
In such a case, owing to the commutation relation, there are more terms in $H_c$ than in $H_p$. Thus, the condition $E(\tau^3H_c)<E(\tau^2mH_p)$ is determined by both $\tau$ and the error Hamiltonians $H_c$ and $H_p$. If $H_c$ is stronger than $H_p$, $\tau$ should be shorter.
In fact, a concrete condition depends on the exact form of the system's Hamiltonian $H_0$ and the decoupling operators used in a decoupling procedure. This is to be discussed in the following section.

\section{Numerical simulations}
\begin{figure}[tbp]
     \centering
     \includegraphics[width=.48\textwidth]{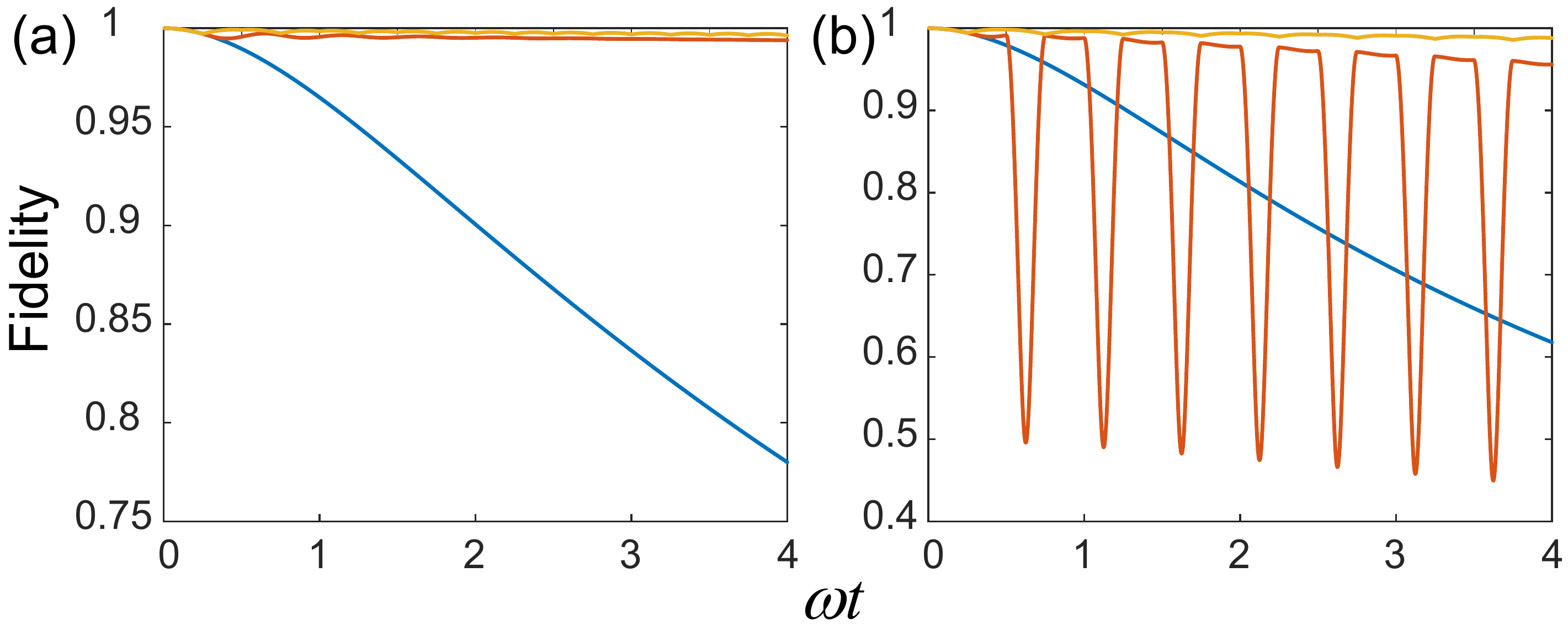}
     \caption{Fidelity between the input $\ket{\psi_k}$ and the corresponding $\rho_k(t)$ versus evolution time. For each input, we consider three cases: (i) with no control Hamiltonian (blue curves), (ii) with the non-ideal control Hamiltonian $J=\pi\omega$ (red curves), and (iii) with the ideal control Hamiltonian $J=\infty$ (yellow curves). (a) Fidelity for $\ket{\psi_1}$. (b) Fidelity for $\ket{\psi_2}$.}
     \label{fig3}
\end{figure}

\begin{figure}[tbh]
  \centering
  \includegraphics[width=.48\textwidth]{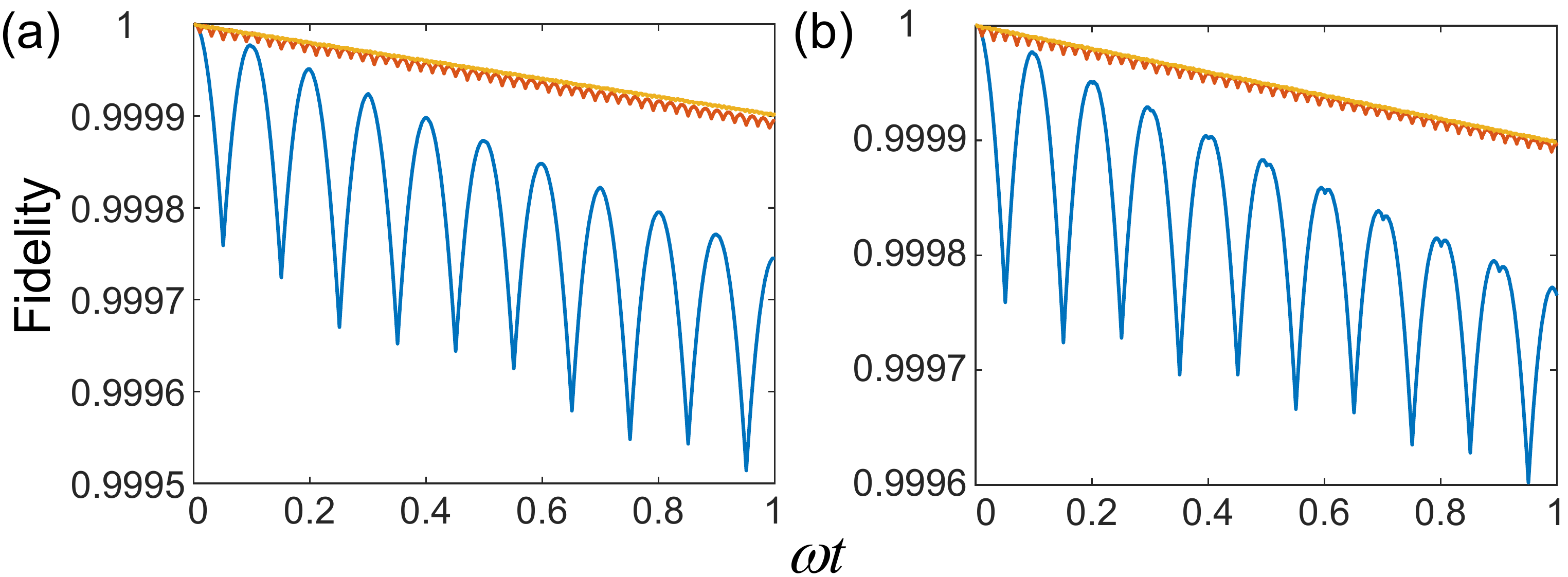}
  \caption{Performance of the concatenated DD in comparison with the periodic DD. We consider three cases: (i) $\tau=\frac{1}{20\omega}$ (blue curves), (ii) $\tau=\frac{1}{100\omega}$ (red curves), (iii) $\tau=\frac{1}{250\omega}$ (yellow curves). (a) Fidelity between the input $\ket{\psi_2}$ and the corresponding $\rho_2(t)$ for the concatenated DD. (b) Fidelity between the input $\ket{\psi_2}$ and the corresponding $\rho_2(t)$ for the periodic DD. }\label{fig4}
\end{figure}

Below we use the quantum Langevin approach~\cite{zhou16,dscontrol} to perform numerical simulations on the DD, so as to
show the validity of our method in more general cases. Here the considered model involves $N$ physical qubits coupled to their bosonic environments independently, as described by
\begin{equation}\label{h0}
  H_{0}=\frac{\omega}{2}\sum_j\sigma_z^j+\sum_{k,j}\omega_k^ja_k^{j\dag}a_k^j+\sum_{k,j}\sigma_{\alpha}^j(g_k^ja_k^j+g_k^{j*}a_k^{j\dag}),
\end{equation}
where $\omega$ is the transition frequency for all qubits and $\sigma^j_{\alpha}$ is the Pauli-$\alpha$ ($\alpha=x,y,z$) operator acting on the $j$th qubit.
Each qubit couples to its environment through $\sigma_{\alpha}^j$ with a coupling strength $|g_k^j|^2$, where $a_k^j$ and $a_k^{j\dag}$ are the annihilation and creation operators of the $k$th bosonic mode with frequency $\omega_k^j$.

For simplicity, all the environments are assumed to be initially in the vacuum state. Thus, the environments can be characterized using their zero-temperature correlation functions $\alpha_{j}(t-s)\equiv\sum_k|g_k^j|^2e^{-i\omega_k^jt}$.
Here we choose the correlation functions of the environments to be the Ornstein-Uhlenbeck type, $\alpha_j(t-s)=\frac{\Gamma_j\gamma_j}{2}e^{-\gamma_j|t-s|}$, where $\Gamma_j=0.1\omega$ denotes the coupling strength between the $j$th qubit and its environment, and $\gamma_j=\omega$ describes the spread of the spectrum \cite{jingthreel,strunzHSPS,jingLEO}.

To achieve the nearest two-qubit state exchange in Fig.~\ref{fig2}, we employ the Heisenberg interaction Hamiltonian as the control Hamiltonian,
\begin{equation}
H_{\rm 1}^{i,j}=J(t)(\sigma_x^i\otimes\sigma_x^{j}+\sigma_y^i\otimes\sigma_y^{j}+\sigma_z^i\otimes\sigma_z^{j}),
\end{equation}
where $J(t)$ describes the ``on" and ``off" of the control Hamiltonian. A state exchange can be realized in a time interval $t_0$ satisfying  $\int_0^{t_0} J(t)\mathrm{d}t=\frac{\pi}{4}$.
The required $N$-qubit state transfer can be constructed using the two-qubit exchange shown in Fig.~\ref{fig2}.

The reduced dynamics of the qubits can be obtained from the non-Markovian quantum Bloch equation~\cite{zhou16},
\begin{equation}
\frac{\partial}{\partial t}\mathcal{A}(t)=-i\mathcal{H}\mathcal{A}(t)+\mathcal{L}\bar{\mathcal{O}}_0(t)\mathcal{A}(t),
\end{equation}
where $\mathcal{A}(t)$ is related to the evolution of the system, $\mathcal{H}$ governs the unitary evolution, and $\mathcal{L}\bar{\mathcal{O}}_0(t)$ generates the non-unitary evolution. The explicit derivation of these operators can be found in \cite{zhou16,dscontrol}.
Given an input $\ket{\psi_k}$, the reduced density matrix $\rho_k(t)$ at time $t$ can be obtained from the non-Markovian quantum Bloch equation.
We use the state fidelity defined as $F=\mathrm{Tr}(\sqrt{\bra{\psi_k}\rho_k(t)\ket{\psi_k}})$ to examine the performance of our method.

We numerically simulate the cases with $N=2$ and $4$ qubits, respectively, each of which comprises a decoherence-free subspace allowing us to encode noise-avoiding states. As shown in Fig.~\ref{fig3}, we use two initial states as inputs: (a) $\ket{\psi_1}$ for the two-qubit case, and (b) $\ket{\psi_2}$ for the four-qubit case.
For each input, we implement three different kinds of numerical simulations: (i) with no control Hamiltonian, (ii) with the non-ideal control Hamiltonian $J=\pi\omega$, and (iii) with the ideal control Hamiltonian $J=\infty$.
The control sequences for the cases of $N=2$ and $4$ directly follow those in Fig.~\ref{fig2}.

Figure~\ref{fig3} presents a clear evidence that the independent noise can be significantly suppressed by our state-transfer-cycle method. For each input, the fidelity between the initial state and its corresponding density matrix at time $t$ decreases rapidly with time in the absence of the control Hamiltonian (blue curves in Fig.~\ref{fig3}), but a modest control Hamiltonian $(J=\pi\omega)$ can greatly suppress this decoherence effect (red curves in Fig.~\ref{fig3}). The achieved fidelities for both the two- and four-qubit cases are greater than 0.95 at the end of the DD procedure ($\omega t=4$).
In the presence of the ideal pulse applied with the same period $\tau=\frac{\pi}{4J}$, the fidelity is improved to be even higher.
Note that for $\ket{\psi_1}$, the fidelities using the ideal and non-ideal pulses almost coincide and are very close to 1. This indicates that a control Hamiltonian with a practical coupling strength is sufficient to implement our scheme.

\begin{table}[tb]
  \centering
\setlength{\tabcolsep}{5mm}{
  \begin{tabular}{|c|c|c|}
     \hline
      & Concatenated DD & Periodic DD \\ \hline
     $\tau=\frac{1}{20\omega}$ & 0.999745 & 0.999765 \\
     $\tau=\frac{1}{100\omega}$ & 0.999895 & 0.999896 \\
     $\tau=\frac{1}{250\omega}$ & 0.999901 & 0.999900 \\
     \hline
   \end{tabular}}
  \caption{Final fidelities for concatenated and periodic DDs in each case.}\label{tab1}
\end{table}

\begin{figure}[tbh]
  \centering
  \includegraphics[width=0.48\textwidth]{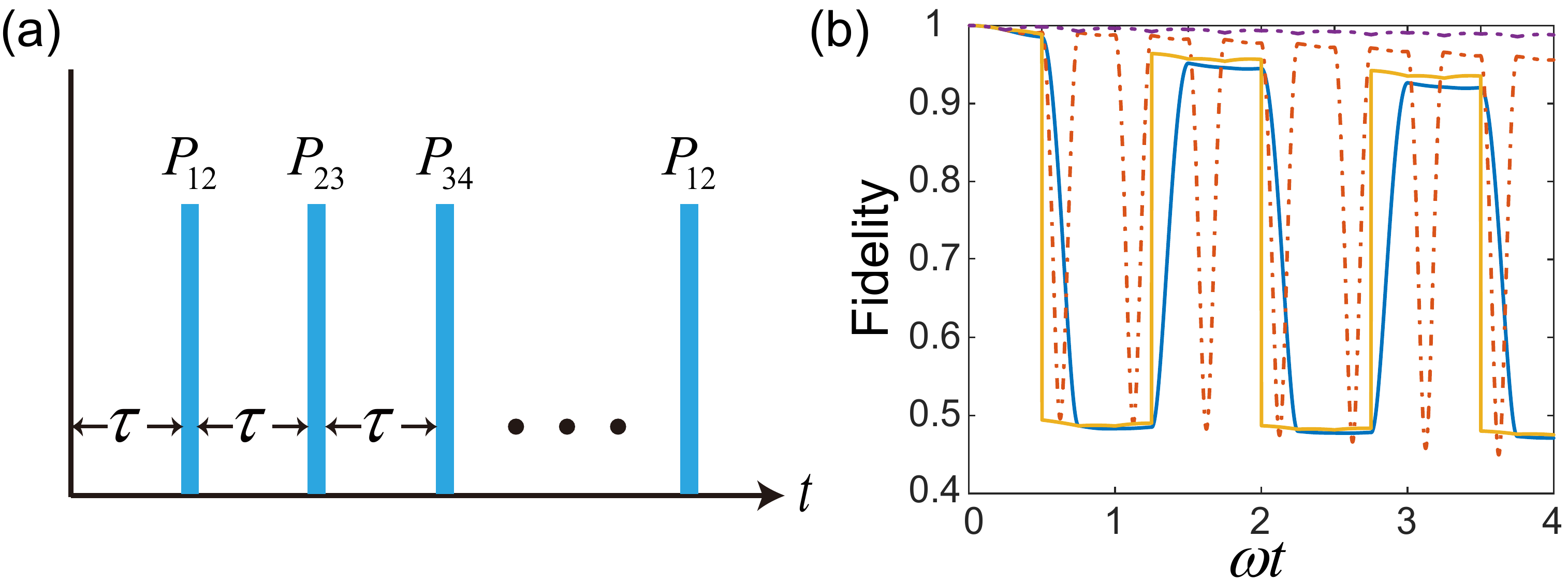}
  \caption{(a) Pulse sequence for the original scheme in \cite{wu02}. (b) Fidelities of the optimal and original schemes. Blue (yellow) curve is for the original scheme with $J=\pi\omega$ ($J=\infty$); red (purple) dashed curve is for the optimal scheme with $J=\pi\omega$ ($J=\infty$).}\label{fig5}
\end{figure}

To compare the performance for both concatenated and periodic DDs, we consider the ideal-pulse case, i.e., the pulses are applied to the qubits instantaneously.
As discussed earlier, when the physical model, i.e., $H_0$ in Eq.~(\ref{h0}), and the decoupling operators $P_1$ and $P_2$ are given, the interval $\tau$ between two adjoining pulses determines the condition $E(\tau^3H_c)<E(\tau^2mH_p)$.
We first choose $\tau=\frac{1}{20\omega}$ to implement the numerical simulation. It turns out that the final fidelity when using the concatenated DD is lower than that using the periodic DD (see Table~\ref{tab1}).
When $\tau=\frac{1}{100\omega}$, the final fidelity when using the concatenated DD is still lower, but almost equal to that using the periodic DD (the difference is less than $1\%$). Reducing $\tau$ further to $\frac{1}{250\omega}$, the final fidelity with the concatenated DD begins to be higher than that using the periodic DD.
Therefore, we can conclude that if we can control the time interval to be $\tau<\frac{1}{250\omega}$, the concatenated DD can be superior to the periodic DD.

Finally, we show the difference between our optimal scheme and the one~\cite{wu02} originally proposed to realize the multi-qubit state transfer operator $P_0$ in Eq.~(\ref{uc}) with a sequence of tow-qubit state exchanges. When $N=4$, the scheme in \cite{wu02} suggests to separate the state-transfer operator $P_0$ to $E_{12}E_{23}E_{34}$, where $E_{ij}$ is the state exchange between the $i$th and $j$th qubits. Note that $H_1^{1,2}$ and $H_1^{2,3}$ are not commutative with each other, so we have to implement the three $E_{ij}$ one by one. Hence, a total of 9 pulses are needed to realize one state-transfer cycle. The distinct advantage of our optimal scheme is that we can instead use $P_1=E_{12}E_{34}$ and $P_{2}=E_{23}E_{14}$, which implies that we can apply two state exchanges at the same time because $H^{1,2}_1$ and $H^{3,4}_1$ commute with each other. In Fig.~\ref{fig5}, we compare the fidelities when using our optimal scheme and the one in \cite{wu02}. It is shown that for the case with $J=\omega\pi$ and the ideal case with $J=\infty$, the final fidelities when using our optimal scheme are higher than those using the original scheme in \cite{wu02}. This demonstrates the efficiency of our scheme.

\section{Discussions and conclusions}
Below we further discuss the underlying mechanism of our approach. We start with the independent two-qubit interaction Hamiltonian
$H_{SB}=\sum_{\alpha} [ \sigma_{\alpha}^{(1)}\otimes B_{\alpha}^{(1)}+ \sigma_{\alpha}^{(2)}\otimes B_{\alpha}^{(2)}]$.
It is shown in Ref.~\cite{wu02} that $H_{SB}$ can be separated into the sum of a collective component $H_{SB}^c=\sum_\alpha[\sigma_\alpha^{(1)}+\sigma_\alpha^{(2)}]\otimes B_\alpha^+$ and a non-collective component $H_{SB}^n=\sum_\alpha[\sigma_\alpha^{(1)}-\sigma_\alpha^{(2)}]\otimes B_\alpha^-$, where $B_\alpha^+=[B_\alpha^{(1)}+B_\alpha^{(2)}]/2$ and $B_\alpha^-=[B_\alpha^{(1)}-B_\alpha^{(2)}]/2$.
We generalize this decomposition for an array of $N$ qubits. The interaction Hamiltonian $H_{SB}$ in Eq.~(\ref{H-sb}) can be rewritten as
\begin{eqnarray}\label{cn}
  H_{SB}&=&\sum_\alpha \left[\sigma_\alpha^{(1)}+\sigma_\alpha^{(2)}+\cdots+\sigma_\alpha^{(N)}\right]\otimes B_\alpha^{1+} \nonumber \\
         &&+\left[\sigma_\alpha^{(1)}-\sigma_\alpha^{(2)}+\cdots+\sigma_\alpha^{(N)}\right]\otimes B_\alpha^{2-} \nonumber \\
         &&+\left[\sigma_\alpha^{(1)}+\sigma_\alpha^{(2)}-\sigma_\alpha^{(3)}+\cdots+\sigma_\alpha^{(N)}\right]\otimes B_\alpha^{3-}
           +\cdots \nonumber \\
         &&+\left[\sigma_\alpha^{(1)}+\sigma_\alpha^{(2)}+\cdots-\sigma_\alpha^{(N)}\right]\otimes B_\alpha^{N-},
\end{eqnarray}
where
\begin{eqnarray}
  B_\alpha^{1+}\! & = &\!\frac{1}{2}\left[B_\alpha^{(2)}+B_\alpha^{(3)}+\cdots+B_\alpha^{(N)}\right]-\frac{N-3}{2}B_\alpha^{(1)}, \nonumber \\
  B_\alpha^{2-}\! & = &\!\frac{1}{2}\left[B_\alpha^{(1)}-B_\alpha^{(2)}\right],\cdots, B_\alpha^{i-}\!=\!\frac{1}{2}\left[B_\alpha^{(1)}-B_\alpha^{(i)}\right],\cdots, \nonumber\\
  B_\alpha^{N-}\! & = &\!\frac{1}{2}\left[B_\alpha^{(1)}-B_\alpha^{(N)}\right].
\end{eqnarray}
It is clear that $B_\alpha^{1+}$ corresponds to the collective component and $B_{\alpha}^{i-}$ ($i=2,3,\cdots,N$) correspond to the non-collective components. While the collective component is invariant under the application of the permutation operators $P_i$ ($i=0,1,2$), $P_i$ will convert the term $-\sigma_\alpha^{(j)}\otimes B_\alpha^{j-}$ in a non-collective component to $-\sigma_\alpha^{(j\pm1)}\otimes B_\alpha^{j-}$, where $\pm$ depends on both $i$ and $j$.
Applying all the operators in $V(t)$ to the $j$th non-collective component, we find that each term in this component gets a negative sign and the sum of all terms becomes a collective component (for details, see Appendix~\ref{c}).

In summary, we have developed a method to create DFSs from the independent error model by using state-transfer cycles of qubits. In our method, the implementation of the state-transfer cycles is shown to be optimal and requires only nearest-neighbor state transfers that can be made with the Heisenberg interaction.
Our scheme makes the strategy in previous approaches~\cite{zanardi99pla,viola00} experimentally feasible by separating the multi-qubit permutations into experimentally realizable two-qubit state transfers. Moreover, our scheme needs only $N$ steps when using two-qubit state transfers to produce an effective collective interaction Hamiltonian, while the previous approaches require at least $(N-1)^2$ steps to generate the same collective interaction Hamiltonian. This polynomial speedup can be significantly important when the number of qubits in the system becomes large to create a higher-dimensional DFS to encode more protected logical qubits for fault-tolerant QC.

For a long period of decoupling, our scheme can be combined with the concatenated DD to suppress the high-order errors. When the time interval between two adjoining pulses is finite, we give a condition under which the concatenated DD is superior to the periodic DD. Also, an explicit time interval for the four-qubit case is found numerically. Our simulations verify the efficiency of our method for the control Hamiltonians with practical coupling strengths.

\begin{acknowledgments}
This work is supported by the National Natural Science Foundation of China (Grant No.~11774022 and U1801661), the National Key Research and Development Program of China (Grant Nos.~2016YFA0301200), and the China Postdoctoral Science Foundation (Grant No.~2018M631437).
Z.-Y. Z. is supported by the Japan Society for the Promotion of Science (JSPS) Foreign Postdoctoral Fellowship No.~P17821.
L.A.W. was supported by the Basque Government (Grant No.~IT986-16) and the Spanish MINECO/FEDER, UE (Grant No.~FIS2015-67161-P).
\end{acknowledgments}

\appendix
\section{The effective Hamiltonian $H_{SB}^{\mathrm{eff}}(T)$ obtained using $V(t)$ in Eq.~(\ref{u1p})}\label{a}
\begin{widetext}
When $V(t)$ in Eq.~(\ref{u1p}) is used as the controller, the corresponding effective Hamiltonian can be written as
\begin{align}
   H_{SB}^{\mathrm{eff}}(T) & =\frac{1}{T_c}\left\{\frac{T_c}{N}H_{SB}+\frac{T_c}{N}P_1^\dag H_{SB}P_1+\frac{T_c}{N}(P_2P_1)^\dag H_{SB}(P_2P_1)+\cdots+\frac{T_c}{N}\left[P_1(P_2P_1)^{\frac{N}{2}-1}\right]^\dag H_{SB}\left[P_1(P_2P_1)^{\frac{N}{2}-1}\right]\right\} \nonumber\\
   & =\frac{1}{N}\left\{H_{SB}+P_1^\dag H_{SB}P_1+(P_2P_1)^\dag H_{SB}(P_2P_1)+\cdots+\left[P_1(P_2P_1)^{\frac{N}{2}-1}\right]^\dag H_{SB}\left[P_1(P_2P_1)^{\frac{N}{2}-1}\right]\right\} \label{total}.
\end{align}
The first term in Eq.~(\ref{total}) is the original interaction Hamiltonian $H_{SB}$,
\begin{equation}\label{exp1}
  H_{SB}=\sum_{\alpha} \left[\sigma_\alpha^{(1)}\otimes B_\alpha^{(1)}+\sigma_\alpha^{(2)}\otimes B_\alpha^{(2)}+\sigma_\alpha^{(3)}\otimes B_\alpha^{(3)}+\sigma_\alpha^{(4)}\otimes B_\alpha^{(4)}+\cdots+\sigma_\alpha^{(N-1)}\otimes B_\alpha^{(N-1)}+\sigma_\alpha^{(N)}\otimes B_\alpha^{(N)}\right].
\end{equation}
The second term in Eq.~(\ref{total}) can be written as
\begin{equation}
  P_1^\dag H_{SB}P_1=\sum_{\alpha} \left[\sigma_\alpha^{(2)}\otimes B_\alpha^{(1)}+\sigma_\alpha^{(1)}\otimes B_\alpha^{(2)}+\sigma_\alpha^{(4)}\otimes B_\alpha^{(3)}+\sigma_\alpha^{(3)}\otimes B_\alpha^{(4)}+\cdots+\sigma_\alpha^{(N)}\otimes B_\alpha^{(N-1)}+\sigma_\alpha^{(N-1)}\otimes B_\alpha^{(N)}\right].
\end{equation}
The third term in Eq.~(\ref{total}) can be written as
  \begin{eqnarray}
  (P_2P_1)^\dag H_{SB}(P_2P_1)&=&\sum_{\alpha} \left[\sigma_\alpha^{(N-1)}\otimes B_\alpha^{(1)}+\sigma_\alpha^{(4)}\otimes B_\alpha^{(2)}+\sigma_\alpha^{(1)}\otimes B_\alpha^{(3)}+\sigma_\alpha^{(6)}\otimes B_\alpha^{(4)}+\cdots\right. \nonumber\\
  &&\left.+\sigma_\alpha^{(N-3)}\otimes B_\alpha^{(N-1)}+\sigma_\alpha^{(2)}\otimes B_\alpha^{(N)}\right].
\end{eqnarray}
Similarly, we can obtain other terms, and the last term in Eq.~(\ref{total}) can be written as
  \begin{eqnarray}\label{expN}
\left[P_1(P_2P_1)^{\frac{N}{2}-1}\right]^\dag H_{SB}\left[P_1(P_2P_1)^{\frac{N}{2}-1}\right]
&=&\sum_{\alpha} \left[\sigma_\alpha^{(N)}\otimes B_\alpha^{(1)}+\sigma_\alpha^{(3)}\otimes B_\alpha^{(2)}+\sigma_\alpha^{(2)}\otimes B_\alpha^{(3)}+\sigma_\alpha^{(5)}\otimes B_\alpha^{(4)}+\cdots \right.  \nonumber\\
&&\left.+\sigma_\alpha^{(N-2)}\otimes B_\alpha^{(N-1)}+\sigma_\alpha^{(1)}\otimes B_\alpha^{(N)}\right].
\end{eqnarray}
\end{widetext}

It can be seen that in the explicit expressions given in Eqs.~(\ref{exp1})-(\ref{expN}), all the first terms are related to $B_\alpha^{(1)}$ and their sum reads
\begin{eqnarray}
  H_{SB}^{\mathrm{eff},1}(T)&=& \left[\sigma_\alpha^{(1)}+\sigma_\alpha^{(2)}+\sigma_\alpha^{(3)}+\cdots+\sigma_\alpha^{(N)}\right]\otimes B_\alpha^{(1)} \nonumber \\
&=&S_\alpha\otimes B_\alpha^{(1)}.
\end{eqnarray}
Similarly, the sum of all the $i$th terms in Eqs.~(\ref{exp1})-(\ref{expN}) can be written as
\begin{eqnarray}
  H_{SB}^{\mathrm{eff},i}(T)&=& \left[\sigma_\alpha^{(1)}+\sigma_\alpha^{(2)}+\sigma_\alpha^{(3)}+\cdots+\sigma_\alpha^{(N)}\right]\otimes B_\alpha^{(i)} \nonumber \\
&=&S_\alpha\otimes B_\alpha^{(i)}.
\end{eqnarray}
Therefore, we have
\begin{eqnarray}\label{21}
  H_{SB}^{\mathrm{eff}}(T)&=&\frac{1}{N}\sum_{\alpha}\sum_{i=1}^{N}H_{SB}^{\mathrm{eff},i}(T) \nonumber \\
               &=& \frac{1}{N}\sum_{\alpha}S_\alpha\otimes\left[B_\alpha^{(1)}+B_\alpha^{(2)}+B_\alpha^{(3)}\cdots+B_\alpha^{(N)}\right] \nonumber \\
&=&\sum_{\alpha}S_\alpha\otimes B_\alpha^{(\rm{env})}.
\end{eqnarray}
Comparing Eq.~(\ref{21}) with Eq.~(\ref{hsbe}), we can conclude that the controller $V(t)$ in Eq.~(\ref{u1p}) produces the same effective Hamiltonian as the controller in Eq.~(\ref{uc}).

\section{Error terms for both periodic and concatenated DDs}\label{b}
To obtain the evolution operator $U_0(T)$ in Eq.~(\ref{u02}), we need to use the BCH formula up to the second order, i.e.,
\begin{equation}\label{}
  e^{Z_1+Z_2}=e^Xe^Y,
\end{equation}
where $Z_1=X+Y$ is the first-order (linear) contribution and $Z_2=\frac{1}{2}[X,Y]$ is the second-order contribution.
Based on this formula, $U_0(T)$ in Eq.~(\ref{u01}) can be written as
\begin{widetext}
\begin{align}\label{}
  U_0(T) & =\prod_{k=0}^{m-1} g_{k}^\dag\exp\{-iH_0\tau\}g_{k}=\prod_{k=2}^{m-1} g_{k}^\dag\exp\{-iH_0\tau\}g_{k}g_{1}^\dag\exp\{-iH_0\tau\}g_{1}\exp\{-ig_0^\dag H_0\tau g_0\} \nonumber\\
   & =\prod_{k=2}^{m-1} g_{k}^\dag\exp\{-iH_0\tau\}g_{k}\exp\{-i\tau(g_1^\dag H_0g_1+g_0^\dag H_0g_0)-\frac{\tau^2}{2}[g_1^\dag H_0g_1,g_0^\dag H_0g_0]\}\nonumber\\
   & =\prod_{k=3}^{m-1} g_{k}^\dag\exp\{-iH_0\tau\}g_{k}g_{2}^\dag\exp\{-iH_0\tau\}g_{2}\exp\{-i\tau(g_1^\dag H_0g_1+g_0^\dag H_0g_0)-\frac{\tau^2}{2}[g_1^\dag H_0g_1,g_0^\dag H_0g_0]\}\nonumber\\
   & =\prod_{k=3}^{m-1} g_{k}^\dag\exp\{-iH_0\tau\}g_{k}\exp\{-i\tau(\sum_{k=0}^{2}
g_k^\dag H_0g_k)-\frac{\tau^2}{2}([g_2^\dag H_0g_2,g_1^\dag H_0g_1]+[g_2^\dag H_0g_2,g_0^\dag H_0g_0]+[g_1^\dag H_0g_1,g_0^\dag H_0g_0])\}\nonumber\\
   & = \cdots \nonumber \\
   & =\exp\{-i\tau\sum_{k=0}^{m-1}g_k^\dag H_0g_k-\frac{\tau^2}{2}\sum_{j>k}[g_j^\dag H_0g_j,g_k^\dag H_0g_k]\}
\end{align}

When the above procedure is implemented successively for $n$ times, the total evolution time is $nT$, and the corresponding
evolution operator in this periodic DD reads
\begin{equation}\label{}
  U_p(nT)=(U_0(T))^n=\exp\{-in\tau\sum_{k=0}^{m-1}g_k^\dag H_0g_k-n\frac{\tau^2}{2}\sum_{j>k}[g_j^\dag H_0g_j,g_k^\dag H_0g_k]\},
\end{equation}
since $U_0(T)$ commutes with itself. When the concatenated DD is employed to implement the DD for a time interval $t=mT$, the total evolution operator can be written as
\begin{align}\label{}
  U_c(mT)=&\prod_{k=0}^{m-1}g_k^\dag U_0(T)g_k=\prod_{k=0}^{m-1}\exp\{-i\tau\bar{H}_0-\frac{\tau^2}{2}\sum_{j>l}g_k^\dag[g_j^\dag H_0g_j,g_l^\dag H_0g_l]g_k\}\nonumber \\
         =&\prod_{k=2}^{m-1}g_k^\dag U_0(T)g_k\exp\{-i\tau\bar{H}_0-\frac{\tau^2}{2}\sum_{j>l}g_1^\dag[g_j^\dag H_0g_j,g_l^\dag H_0g_l]g_1\}\exp\{-i\tau\bar{H}_0-\frac{\tau^2}{2}\sum_{j>l}g_0^\dag[g_j^\dag H_0g_j,g_l^\dag H_0g_l]g_0\} \nonumber \\
         =&\prod_{k=2}^{m-1}g_k^\dag U_0(T)g_k\exp\{-i2\tau\bar{H}_0-\frac{\tau^2}{2}(g_1^\dag[j,l]g_1+g_0^\dag[j,l]g_0)+i\frac{\tau^3}{4}([\bar{H}_0,g_0^\dag[j,l]g_0]+[g_1^\dag[j,l]g_1,\bar{H}_0])\}\nonumber \\
  =&\prod_{k=3}^{m-1}g_k^\dag U_0(T)g_k\exp\{-i3\tau\bar{H}_0-\frac{\tau^2}{2}\sum_{k=0}^{2}g_k^\dag[j,l]g_k+i\frac{\tau^3}{4}[\bar{H}_0,g_0^\dag[j,l]g_0-2g_2^\dag[j,l]g_2]\} \nonumber \\
  =&\cdots \nonumber \\
  =&\exp\{-im\tau\bar{H}_0-\frac{\tau^2}{2}\sum_{k=0}^{m-1}g_k^\dag[j,l]g_k+i\frac{\tau^3}{4}\sum_{k=0}^{m-1}(m-2k-1)[\bar{H}_0,g_k^\dag[j,l]g_k]\},
\end{align}
where $\bar{H}_0=\sum_{k=0}^{m-1}g_k^\dag H_0g_k$, and $[j,l]=\sum_{j>l}[g_j^\dag H_0g_j,g_l^\dag H_0g_l]$.
\end{widetext}

\section{Elimination of the non-collective components in Eq.~(\ref{cn})}\label{c}
To produce a collective Hamiltonian $H_{SB}^\mathrm{eff}$, we should eliminate all the non-collective components in $H_{SB}$.
Consider the $j$th component in Eq.~(\ref{cn}),
\begin{eqnarray}\label{hj}
H_{SB}^j&=&\left[\sigma_\alpha^{(1)}+\sigma_\alpha^{(2)}+\cdots-\sigma_\alpha^{(j)}+\sigma_\alpha^{(j+1)}
+\cdots+\sigma_\alpha^{(N)}\right] \nonumber\\
 &&\otimes B_\alpha^{j-}.
\end{eqnarray}
The permutation operators $P_i$ ($i=0,1,2$) in both Eq.~(\ref{uc}) and Eq.~(\ref{u1p}) can convert a term $\sigma_\alpha^{(k)}\otimes B_\alpha^{j-}$ in Eq.~(\ref{hj}) to $\sigma_\alpha^{(k\pm1)}\otimes B_\alpha^{j-}$.
Explicitly, we have
\begin{align}\label{}
   & P_0^\dag \sigma_\alpha^{(k)}\otimes B_\alpha^{j-}P_0=\sigma_\alpha^{(k-1)}\otimes B_\alpha^{j-}  \nonumber\\
   & P_1^\dag \sigma_\alpha^{(k)}\otimes B_\alpha^{j-}P_1=\left\{
       \begin{array}{ll}
        \sigma_\alpha^{(k+1)}\otimes B_\alpha^{j-} , & \hbox{$k=$~even;} \\
        \sigma_\alpha^{(k-1)}\otimes B_\alpha^{j-}, & \hbox{$k=$~odd.}
       \end{array}
     \right. \\
   & P_2^\dag \sigma_\alpha^{(k)}\otimes B_\alpha^{j-}P_2=\left\{
       \begin{array}{ll}
        \sigma_\alpha^{(k+1)}\otimes B_\alpha^{j-} , & \hbox{$k=$~even;} \nonumber\\
        \sigma_\alpha^{(k-1)}\otimes B_\alpha^{j-}, & \hbox{$k=$~odd.}
       \end{array}
     \right.
\end{align}
Let us take $V(t)$ in Eq.~(\ref{uc}) as an example.
When $P_0$ is applied to $H_{SB}^j$, it gives rise to a new term
\begin{equation}
H_{SB}^{j,1}=\left[\sigma_\alpha^{(1)}+\cdots-\sigma_\alpha^{(j-1)}+\sigma_\alpha^{(j)}\cdots+\sigma_\alpha^{(N)}\right]\otimes B_\alpha^{j-}.
\end{equation}
Applying $P_0$ twice (i.e., $P_0^2$), we have
\begin{equation}\label{}
  H_{SB}^{j,2}=\left[\sigma_\alpha^{(1)}+\cdots-\sigma_\alpha^{(j-2)}+\sigma_\alpha^{(j-1)}\cdots+\sigma_\alpha^{(N)}\right]\otimes B_\alpha^{j-}.
\end{equation}
Similarly, the term $H_{SB}^{j,m}$ can be obtained by applying $P_0$ $m$ times.
Then, summing up the original term $H_{SB}^j$ and all the terms $H_{SB}^{j,m}$, $m=1$ to $N-1$, we obtain a collective component
\begin{align}\label{}
  H_{SB}^{j,c}=&H_{SB}^j+\sum_{m=1}^{N-1}H_{SB}^{j,m} \nonumber \\
  =&(N-1)\left[\sigma_\alpha^{(1)}+\sigma_\alpha^{(2)}+\cdots+\sigma_\alpha^{(N)}\right]\otimes B_\alpha^{j-}.
\end{align}
Because we consider an arbitrary (i.e., the $j$th) term in the above derivations, it follows that all the non-collective components are then eliminated simultaneously and finally we obtain a collective effective Hamiltonian.

\end{document}